\documentclass[a4paper]{spie}  

\usepackage{amsmath,amsfonts,amssymb}
\usepackage{graphicx}
\usepackage[colorlinks=true, allcolors=blue]{hyperref}

\title{Ultrafast spatiotemporal dynamics of a charge-density wave using femtosecond dark-field momentum microscopy}

\author[a]{J. Maklar}
\author[b, c]{P. Walmsley}
\author[b, c]{I.R. Fisher}
\author[a]{L. Rettig}
\affil[a]{Fritz-Haber-Institut der Max-Planck-Gesellschaft, Faradayweg 4-6, 14195 Berlin, Germany}
\affil[b]{Geballe Laboratory for Advanced Materials and Department of Applied Physics, Stanford University, Stanford, CA, USA}
\affil[c]{Stanford Institute for Materials and Energy Sciences, SLAC National Accelerator Laboratory, Menlo Park, CA, USA}

\authorinfo{Further author information: (Send correspondence to L.R.)\\L.R.: E-mail: rettig@fhi-berlin.mpg.de}

\begin{document} 
\maketitle

\begin{abstract}
Understanding phase competition and phase separation in quantum materials requires access to the spatiotemporal dynamics of electronic ordering phenomena on a micro- to nanometer length- and femtosecond timescale. While time- and angle-resolved photoemission (trARPES) experiments provide  sensitivity to the femtosecond dynamics of electronic ordering, they typically lack the required spatial resolution. Here, we demonstrate ultrafast dark-field photoemission microscopy (PEEM) using a momentum microscope, providing access to ultrafast electronic order on the microscale. We investigate the prototypical charge-density wave (CDW) compound TbTe$_3$ in the vicinity of a buried crystal defect, demonstrating real- and reciprocal-space configurations combined with a pump-probe approach. We find CDW order to be suppressed in the region covered by the crystal defect, most likely due to locally imposed strain. Comparing the ultrafast dynamics in different areas of the sample reveals a substantially smaller response to optical excitation and faster relaxation of excited carriers in the defect area, which we attribute to enhanced particle-hole scattering and defect-induced relaxation channels.
\end{abstract}

\keywords{Time- and angle-resolved photoemission, charge density wave dynamics, femtosecond dark-field photoemission microscopy}

\section{INTRODUCTION}
\label{sec:intro}
Quantum materials, i.e. systems whose macroscopic properties are governed by quantum many-body effects, promise fascinating future applications in material science. Their properties are governed by complex interactions between elementary degrees of freedom -- electronic, spin, orbital, and lattice -- giving rise to fascinating collective phenomena such as unconventional, high-temperature superconductivity, quantum spin liquids, Mott transitions, or topological states\cite{Keimer2017_NatPhys, Tokura2017_NatPhys, Sobota2021_RevModPhys_trARPES, Narang2021_NatMater}. A particularly ubiquitous collective ground state found in many materials are so-called charge density waves (CDWs), i.e. a symmetry-broken phase characterized by a charge superstructure coupled to a periodic lattice deformation\cite{gruner1994}. CDW order is often found in close proximity to other correlated phases such as superconductivity\cite{Lee2006} or magnetic order\cite{Li2022}. This proximity frequently leads to phase coexistence and inhomogeneities on the micro- to nanoscale, which have been suggested to play a decisive role for some of their fascinating properties\cite{Campi2015}.

An established approach to gain a better understanding of quantum materials is to study their dynamical response to a tailored ultrashort optical excitation. While in thermal equilibrium, various interactions between elementary degrees of freedom are present simultaneously, and their specific coupling often cannot be accessed due to a lack of directly measurable observables, time-resolved methods overcome this problem by investigating interactions directly in the time domain on the intrinsic timescale of fundamental scattering processes. A powerful method for such investigations is time- and angle-resolved photoemission spectroscopy (trARPES), which provides direct sensitivity to the transient electronic band structure, and thus the dynamics of electronic ordering processes, e.g. in CDW materials\cite{schmitt2008, perfetti2006, hellmann2012time, Maklar2021_NatCommun} or superconductors\cite{perfetti2007_PRL_cuprate, Cortes2011_PRL_cuprate_dynamics, Smallwood2012_BSCCO}. However, understanding the interplay of coexisting phases on the micro- and nanoscale requires methods capable of resolving the respective length scales, in addition to femtosecond temporal resolution. While for studying structural dynamics on the nanoscale methods such as ultrafast dark-field electron microscopy\cite{Danz2021_nanoimaging} or femtosecond x-ray holography\cite{Johnson2022} have been demonstrated, typical trARPES experiments average the dynamics over the spot size of the probe pulse of typically several 10s of micrometer.

\begin{figure} [tb]
\centering
\includegraphics[width=1\textwidth]{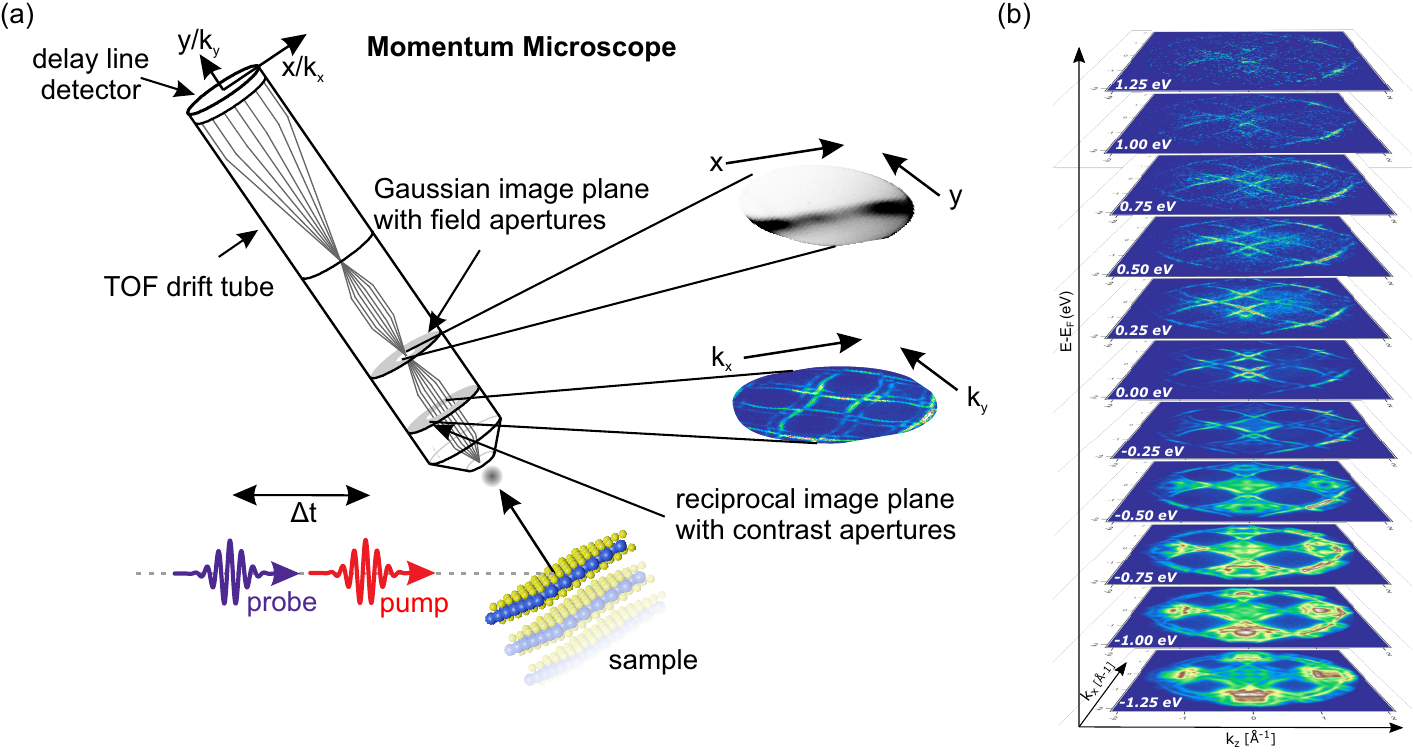}
\caption{\textbf{Experimental scheme and band structure mapping} (a) Schematic of the trARPES experiment and momentum microscope. In the electrostatic lens system, intermediate real and reciprocal space images are formed, which can be filtered using field and contrast apertures, respectively. (b) Time-resolved ARPES data of TbTe$_3$ at temporal pump-probe overlap for various energies below and above E$_F$ at an absorbed fluence of $F_{\mathrm{abs}}=30$\,\textmu J\,cm$^{-2}$.}
\label{fig:experiment} 
\end{figure}

The advent of a novel type of electron analyzers, so-called Momentum Microscopes\cite{Schonhense2015_Elsevier}, enable a similar approach for investigating the ultrafast dynamics of the electronic structure. As sketched in Fig.~\ref{fig:experiment}(a), the momentum microscope contains sophisticated photoemission microscopy (PEEM) electron optics, which generate virtual images of the real- and momentum-space distribution of photoemitted electrons. These virtual images can be selectively projected onto a position-sensitive delay line detector, which encodes the electron energy in the arrival time of each electron on the detector after passing a time-of-flight (TOF) section. The use of field apertures in the real-space image plane permits spatial selectivity in band structure measurements down to the \textmu m scale, while the use of contrast apertures in the reciprocal image plane enables dark-field PEEM, i.e. momentum selectivity for measurements of the real-space photoelectron distribution. Combining these approaches with ultrashort laser pulses and a pump-probe scheme provides access to the dynamics of spatial inhomogeneities on the order of 10s of fs, at few \textmu m spatial resolution, $\sim 150$\,meV energy resolution, and at high momentum selectivity. 

Here, we present trARPES experiments on the prototypical CDW compound TbTe$_3$ in the vicinity of a buried crystal defect, investigating the femtosecond dynamics of CDW ordering in both real- and reciprocal-space configurations. Using area-selective momentum microscopy allows us to selectively study the electronic band structure within the defect region, demonstrating the absence of CDW order in this area. Momentum selective dark-field PEEM further allows us to investigate the spatiotemporal dynamics of the response to photoexcitation, revealing suppressed excitation and faster quasiparticle relaxation in the defect region.

\section{EXPERIMENT}
\label{sec:exp}

trARPES measurements have been performed on single crystals of TbTe$_3$ as described earlier\cite{Maklar2021_NatCommun}. The setup includes a tabletop fs XUV source ($h\nu_\mathrm{probe} = 21.7\,$eV) with a synchronized optical pump laser ($h\nu_\mathrm{pump} = 1.55\,$eV) operating at a repetition rate of 500~kHz\cite{puppin2019_RSI}. Electrons photoemitted by the pump pulse were detected using a momentum microscope (SPECS Metis 1000)\cite{Maklar2020_MetisPhoibos}. The combined time and energy resolutions are $\sim$35~fs and $\sim$150~meV, respectively, and the achievable momentum and spatial resolutions are $\sim0.08\,$\AA$^{-1}$ and $\sim1\,\mu$m. Single crystals of TbTe$_3$ were grown by slow cooling of a binary melt\cite{ru2006}, and cleaved in-situ in ultra-high vacuum $<1\times10^{-10}$ mbar, where measurements have been carried out. The pump and probe spot sizes (FWHM) are $\approx230 \times 200\,\mu$m$^2$ and $\approx70 \times 60 \mu$m$^2$, respectively, and absorbed fluences F$_\mathrm{abs}$ were estimated from the complex refractive n = 0.9 +i2.6\cite{Maklar2021_NatCommun}. All experiments were carried out at $T=80$~K, well below $T_\mathrm{c}=336$~K, the transition temperature of the unidirectional CDW phase\cite{ru2008}.

The momentum microscope in combination with the high repetition rate of the trARPES setup allows for investigating the electronic band structure beyond the first Brillouin zone (BZ), and in a large energy range below and above the Fermi level, by photoemitting electrons transiently excited into unoccupied states\cite{Puppin2022_PRB_bandmapping}. The corresponding momentum-dependent band structure of TbTe$_3$ at pump-probe overlap is shown for various energies in Fig.~\ref{fig:experiment}(b), exhibiting well-resolved states up to 1.5~eV above E$_F$. The electronic properties near E$_F$ are governed by a square lattice of quasi-2D Te sheets, which give rise to a diamond-shaped hole-pocket around the $\Gamma$-Point, dispersing inwards with energy\cite{brouet2008}. 

\begin{figure} [tb]
\centering
\includegraphics[width=1\textwidth]{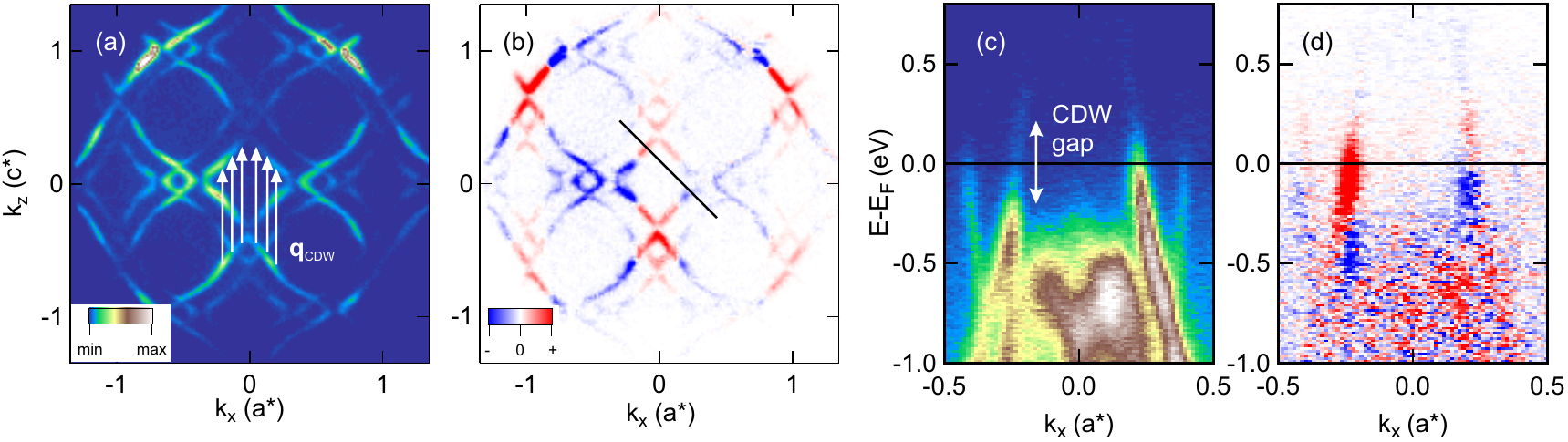}
\caption{\textbf{Time-resolved Fermi surface and gap dynamics} (a) Fermi surface of TbTe$_3$ at negative pump-probe delay. CDW energy gaps appear in FS regions nested by the nesting vector $\mathbf{q}_{CDW}$. (b) Pump-induced differences of the FS intensity at 250~fs. (c) Energy-Momentum dispersion along the black line in panel (b) at temporal pump-probe overlap, showing the CDW energy gap opening at negative $k_x$. (d) Pump-induced differences at $\Delta t=250$~fs along the cut shown in (b). Measurements were taken with an absorbed fluence of $F_{\mathrm{abs}}=30$\,\textmu J\,cm$^{-2}$.}
\label{fig:TR_FS} 
\end{figure} 

For understanding the CDW properties of the pristine band structure, Fig.~\ref{fig:TR_FS}(a) shows the Fermi surface (FS) of TbTe$_3$, also featuring the diamond-shaped form of the band dispersion. Below $T_\mathrm{c}$, strongly wave vector-dependent electron-phonon coupling\cite{maschek2015}, in conjunction with a moderately well-nested Fermi surface\cite{laverock2005fermi} lead to the formation of a unidirectional CDW, in which some portions of the FS that are connected by the nesting vector $\mathbf{q}_{CDW}$ are gapped while others remain metallic\cite{brouet2008} (see Fig.~\ref{fig:TR_FS}(a)). This manifests as momentum-stripes along the $k_z$ direction featuring an energy gap opening at $E_{\mathrm{F}}$. Note that, by convention, the in-plane crystal directions are along the x and z ($k_x$ and $k_z$) axes. After pump excitation, these gapped-out regions on the FS strongly gain in intensity, while the metallic regions lose intensity (Fig.~\ref{fig:TR_FS}(b)). This behavior demonstrates a transient closing of the CDW energy gaps on the timescale of ~200 fs, consistent with literature\cite{schmitt2008, schmitt2011, Maklar2021_NatCommun, Maklar2021_PRL}. The energy-momentum cut along the black line in Fig.~\ref{fig:TR_FS}(b) is shown in panel (c), corroborating the opening of the CDW energy gap around E$_F$ at negative $k_x$, while the band at positive $k_x$ remains metallic. After excitation, the closing of the CDW energy gap manifests as strong increase of the intensity within the CDW energy gap, and transfer of spectral weight out of the metallic band (Fig.~\ref{fig:TR_FS}(d))

\section{RESULTS}
\subsection{Time- and spatially-resolved ARPES.}
\label{sec:PEEM}

\begin{figure}[tb]
\centering
\includegraphics[width=1\textwidth]{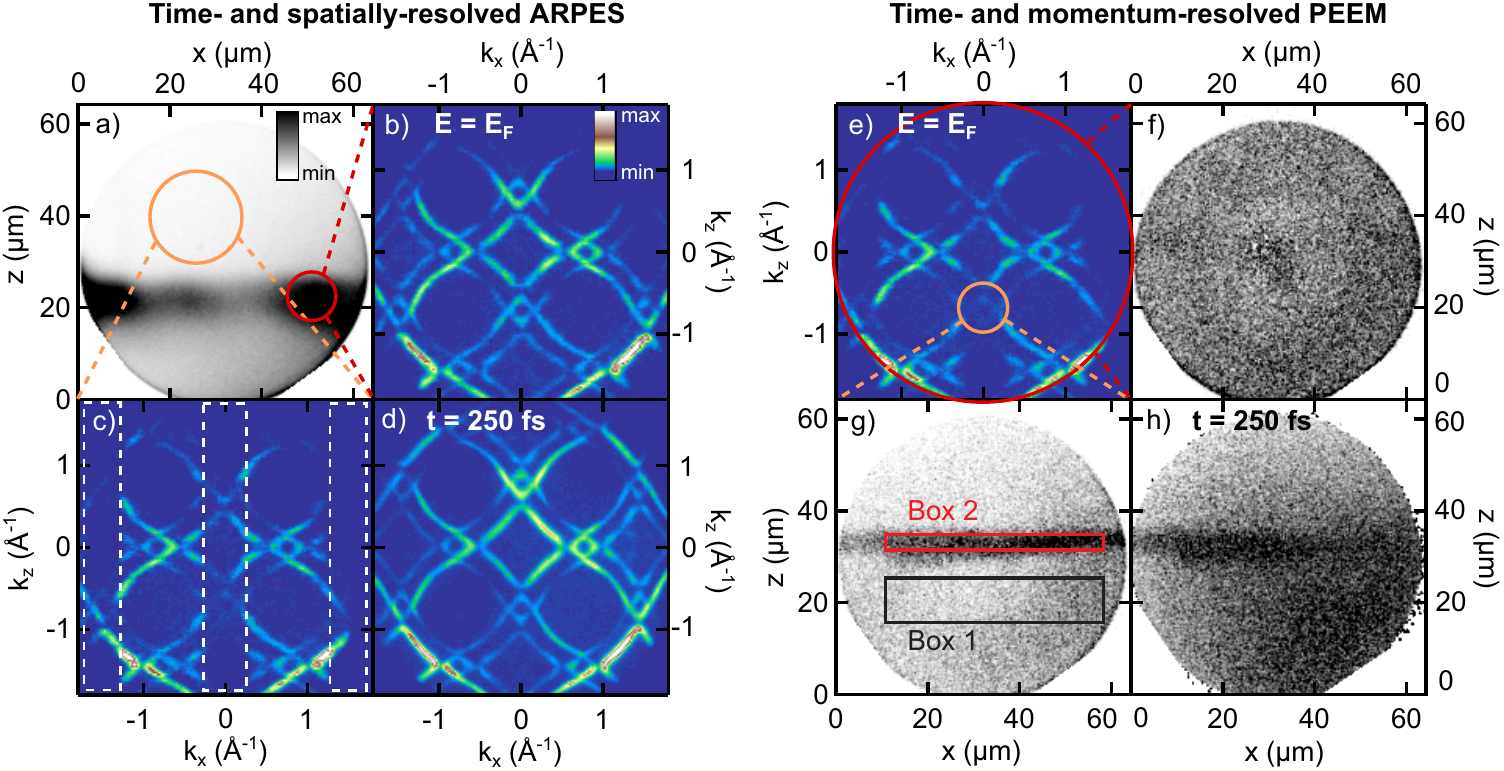}
\caption{\textbf{Selected area ARPES and dark-field PEEM imaging} (a) Energy-integrated PEEM image of pump-laser-induced photoemission signal from a crystal defect in TbTe$_3$ ($h\nu=1.55$\,eV, absorbed fluence $F_{\mathrm{abs}}=140$\,\textmu J\,cm$^{-2}$). (b) Static FS of the defect region of TbTe$_3$ ($E_{\mathrm{F}}\pm50$\,meV, field aperture diameter $d_{\mathrm{FA}}=100$\,\textmu m). (c) Static FS of the defect-free sample area ($d_{\mathrm{FA}}=200$\,\textmu m). White boxes mark the CDW-gapped regions. (d) FS of the defect-free sample region $\Delta t=250$\,fs after photoexcitation. (e) Static FS, same as panel (c). (f) Static PEEM image at $E_{\mathrm{F}}$ of the sample spot characterized in panel (a) without a contrast aperture and (g) with a contrast aperture ($d_{\mathrm{CA}}=200$\,\textmu m) focused on the CDW-gapped area. (h) PEEM image with contrast aperture after photoexcitation ($h\nu_\mathrm{pump}=1.55$\,eV, $F_{\mathrm{abs}}=90$\,\textmu J\,cm$^{-2}$, $\Delta t=250$\,fs).}
\label{fig:PEEM}
\end{figure}

Using the spatially resolved PEEM mode of the momentum microscope, we identify an inhomogeneity on the TbTe$_3$ sample surface. Inhomogeneities such as crystal defects typically reduce the work function and lead to field enhancement of the pump beam by plasmonic effects. Due to the nonlinearity of the multi-photon photoemission signal generated by $1.55\,$eV laser pulses, the reduced work function and field enhancement locally strongly increase the pump photoemission signal, resulting in a stark intensity contrast between defects and pristine sample regions. Fig.~\ref{fig:PEEM}(a) shows the multiphoton PEEM image from the pump beam of a horizontally traversing defect in TbTe$_3$. To investigate the influence of the defect on the electronic band structure and the CDW, we first perform trARPES measurements on the defect-free area above the defect by inserting a suitable field aperture in the real-space image plane (indicated in orange in Fig.~\ref{fig:PEEM}(c)). At equilibrium, we find the characteristic suppression of photoemission intensity at $E_\mathrm{F}$ within the CDW gapped areas (Fig.~\ref{fig:PEEM}(c)). As discussed above, ultrafast optical excitation transiently melts the CDW within a few hundred fs\cite{schmitt2008}, evident from the intensity gain in the previously gapped regions, resulting in a continuous Fermi surface (Fig.~\ref{fig:PEEM}(d)). Interestingly, when selecting photoelectrons originating only from the defect region by inserting a smaller field aperture (indicated in red in Fig.~\ref{fig:PEEM}(a)), we observe a fully metallic, ungapped FS even at equilibrium, (Fig.~\ref{fig:PEEM}(b)), demonstrating suppression of CDW order within the region of the defect. 

\subsection{Ultrafast dark-field PEEM imaging}
\label{sec:dark-field}
Next, we employ the possibility to place contrast apertures in the intermediate momentum-space image plane, to investigate the spatial distribution of the CDW order parameter. This approach is similar to dark-field microscopy e.g. employed in electron microscopy\cite{Wilson1975_TaS2_CDW}, which has recently been demonstrated to allow studying the femtosecond and nanometer dynamics of CDW order\cite{ Danz2021_nanoimaging}. Application of dark-field PEEM has been recently applied to separate differently oriented domains in polycristalline graphene films\cite{Barrett2012, Barrett2013, Wan2020}. Here, we for the first time demonstrate the application of this technique to investigate the ultrafast microscale dynamics of electronically ordered phases in complex materials.

In Fig.~\ref{fig:PEEM}(e)-(h), we present femtosecond time- and micrometer spatially resolved PEEM measurements in the surrounding of the identified defect. First, when no contrast aperture is inserted in the reciprocal image plane, the spatial distribution of photoelectrons emitted by the XUV probe is averaged over all momenta across the photoemission horizon, analogously to a bright-field image in electron microscopy. Since the CDW only partially gaps the FS, also sample regions featuring CDW order host electrons up to $E_{\mathrm{F}}$. Therefore, the real-space electron distribution at $E_{\mathrm{F}}$ in bright-field operation is largely homogeneous, as shown in panel (f). In contrast, when placing a small contrast aperture on the CDW-gapped region of the FS (orange circle in panel (e)), we observe a clear contrast between the defect, which hosts metallic states in the selected energy-momentum region, and the surrounding CDW-gapped area (Fig.~\ref{fig:PEEM}(g)). The clear signal from the defect observed in this dark-field mode of operation demonstrates that the signal at $E_{\mathrm{F}}$ in the selected momentum area originates exclusively from sample regions where CDW formation is suppressed, corresponding to the footprint of the defect. Note, that the signal remains homogeneous for energies below the CDW energy gap. 

\begin{figure} [tb]
\centering
\includegraphics[width=1\textwidth]{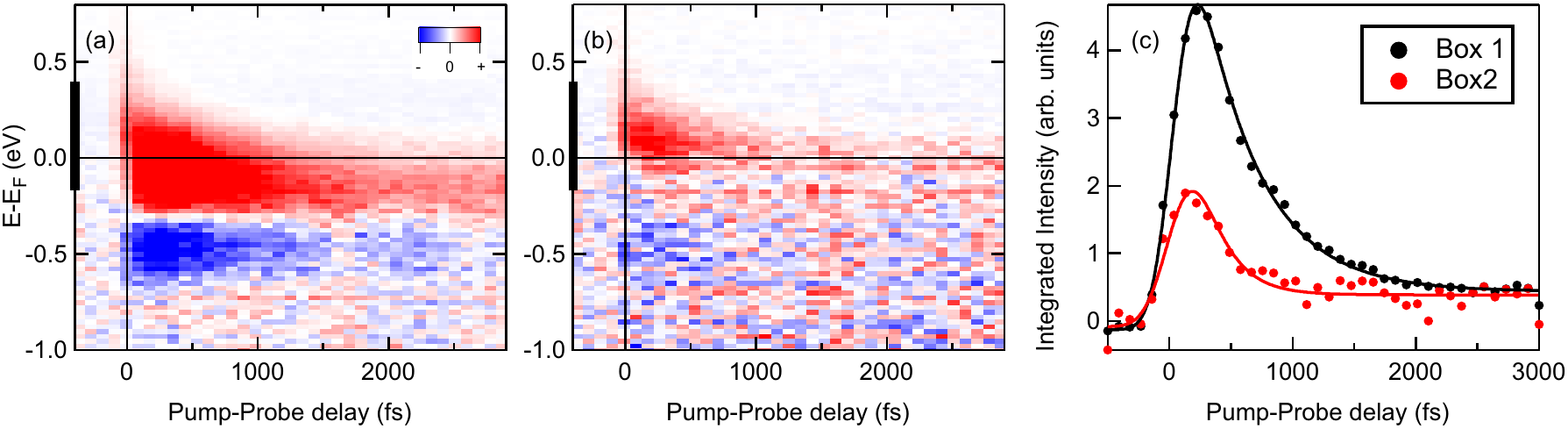}
\caption{\textbf{Spatially resolved relaxation dynamics of excited carriers.} Energy- and pump-probe delay-dependent pump-induced photoemission signal in the CDW-gapped region (Box 1, panel (a)), and from the defect (Box 2, panel (b)), respectively. The transient photoemission signal integrated around the Fermi level (black bars in (a), (b)) is shown in (c), together with monoexponential fits (see text).}
\label{fig:decay_dynamics} 
\end{figure}

This mode of operation allows us to investigate the spatial distribution of the ultrafast dynamics after photoexcitation, and to investigate the respective response in different parts of the sample. For this we combine momentum-selective  dark-field PEEM with a pump-probe approach, revealing a partial suppression of the CDW gap, and a corresponding metallization in sample regions next to the defect, as shown in Fig.~\ref{fig:PEEM}(h). For a more detailed analysis of the ultrafast dynamics in the regions gapped by the CDW and at the sample defect line, we integrate the signal in respective areas, marked as 'Box 1' and 'Box 2' in Fig.~\ref{fig:PEEM}(g). Corresponding traces of the pump-induced photoemission signal as function of energy and pump-probe delay are shown in Fig.~\ref{fig:decay_dynamics}(a) and (b). Remarkably, we observe a much stronger response in the region featuring the CDW energy gap (panel (a)), compared to the area around the defect (panel (b)). In the CDW-gapped region, the strong increase of intensity for energies above $\sim-250\,$meV (red), and decrease of intensity below (blue) corresponds to the closing of the CDW energy gap, and unfolding of the nested band structure, in agreement with the momentum-space picture shown in Fig.~\ref{fig:TR_FS}. In contrast, the area near the defect shows only a small increase of intensity above and minor decrease below E$_F$, corresponding to particle-hole excitations in a trivial metal. Fig.~\ref{fig:decay_dynamics}(c) shows the transient photoemission signal integrated at energies around the Fermi level, as indicated by the black markers in Fig.~\ref{fig:decay_dynamics}(a) and (b). Apart from the reduced signal in the defect region, we also observe substantially longer life times of transient carriers in the CDW-gapped region. To quantify the relaxation dynamics, we fit the curves with a monoexponential decay function, yielding decay times of $480\pm80\,$fs for the CDW-gapped region, and $220\pm80\,$fs for the defect area, respectively.

\section{DISCUSSION}
Our observations bring up the question about the nature of the observed defect. As the band structure in the defect region features sharp, undistorted electronic bands corresponding to the high-temperature state, the crystal defect is likely a crystal dislocation buried below the surface, which imposes strain on the topmost crystal layers and thus suppresses CDW formation, while leaving the underlying band dispersion intact. Indeed, selective suppression of CDW order under application of uniaxial strain has been recently reported in the CDW compound 1$T$-TaS$_2$\cite{Nicholson2022_TaS2}, and a pressure-induced suppression of CDW order and emergence of superconductivity at low temperatures was also observed for various members of the RTe$_3$ family\cite{Zocco2015}. 

Remarkably, even though we see enhanced pump-induced photoemission from the defect area (Fig.~\ref{fig:PEEM}(a)), indicating a field-enhanced cross-section for pump absorption, a smaller response to pump excitation is observed in the photoemission signal. This highlights the enhanced susceptibility to optical excitation in the CDW-state, which drives the transition into the transient metallic state. Furthermore, the faster relaxation in the defect area is intriguing. This observation is in line with enhanced particle-hole scattering rates observed recently in the transient metallic state\cite{Maklar2021_PRL}. Additionally, defect-induced relaxation pathways might further contribute to the quasiparticle relaxation.

\section{CONCLUSIONS}
In conclusion we investigate the spatiotemporal ultrafast dynamics of a CDW-to-metal transition in the vicinity of a crystal defect using femtosecond dark-field PEEM imaging and area-selective momentum microscopy. While the electronic band structure remains intact, the crystal defect is found to suppress CDW order, most likely due to locally induced strain. Time-resolved dark-field PEEM images reveal a substantially small response to optical excitation and faster relaxation of excited carriers in the defect area, which we explain by enhanced particle-hole scattering in the metallic state, and defect-induced relaxation channels. Our findings establish femtosecond dark-field PEEM imaging as a promising technique to study the ultrafast spatiotemporal dynamics of correlated materials on a femtosecond time- and sub-micrometer spatial scale.

\acknowledgments 
This work was funded by the Max Planck Society, and the German Research Foundation (DFG) within the Emmy Noether program (Grant No. RE 3977/1). Crystal growth and characterization at Stanford University (P. W. and I. R. F.) was supported by the Department of Energy, Office of Basic Energy Sciences under Contract No. DE-AC02-76SF00515.

\bibliographystyle{spiebib} 

\begin{thebibliography}{10}

\bibitem{Keimer2017_NatPhys}
Keimer, B. and Moore, J.~E., ``{The physics of quantum materials},'' {\em Nat.
  Phys.}~{\bf 13}(11),  1045--1055 (2017).

\bibitem{Tokura2017_NatPhys}
Tokura, Y., Kawasaki, M., and Nagaosa, N., ``{Emergent functions of quantum
  materials},'' {\em Nat. Phys.}~{\bf 13}(11),  1056--1068 (2017).

\bibitem{Sobota2021_RevModPhys_trARPES}
Sobota, J.~A., He, Y., and Shen, Z.-X., ``Angle-resolved photoemission studies
  of quantum materials,'' {\em Rev. Mod. Phys.}~{\bf 93},  025006 (May 2021).

\bibitem{Narang2021_NatMater}
Narang, P., Garcia, C. A.~C., and Felser, C., ``{The topology of electronic
  band structures},'' {\em Nat. Mater.}~{\bf 20}(3),  293--300 (2021).

\bibitem{gruner1994}
Gruner, G.,  [{\em \href{https://doi.org/10.1201/9780429501012}{Density waves
  in solids}}{\nolinebreak\hspace{0.1em}]}, CRC press (1994).

\bibitem{Lee2006}
Lee, J., Fujita, K., McElroy, K., Slezak, J.~A., Wang, M., Aiura, Y., Bando,
  H., Ishikado, M., Masui, T., Zhu, J.-X., Balatsky, A.~V., Eisaki, H., Uchida,
  S., and Davis, J.~C., ``Interplay of electron-lattice interactions and
  superconductivity in {B}i$_2${S}r$_2${C}a{C}u$_2$o$_{8+\delta}$,'' {\em
  Nature}~{\bf 442}(7102),  546--550 (2006).

\bibitem{Li2022}
Li, J., Gorobtsov, O.~Y., Patel, S. K.~K., Hua, N., Gregory, B., Shabalin,
  A.~G., Hrkac, S., Wingert, J., Cela, D., Glownia, J.~M., Chollet, M., Zhu,
  D., Medapalli, R., Fullerton, E.~E., Shpyrko, O.~G., and Singer, A.,
  ``Phonon-assisted formation of an itinerant electronic density wave,'' {\em
  Commun. Phys.}~{\bf 5}(1),  125 (2022).

\bibitem{Campi2015}
Campi, G., Bianconi, A., Poccia, N., Bianconi, G., Barba, L., Arrighetti, G.,
  Innocenti, D., Karpinski, J., Zhigadlo, N.~D., Kazakov, S.~M., Burghammer,
  M., Zimmermann, M.~v., Sprung, M., and Ricci, A., ``Inhomogeneity of
  charge-density-wave order and quenched disorder in a high-{T}$_c$
  superconductor,'' {\em Nature}~{\bf 525}(7569),  359--362 (2015).

\bibitem{schmitt2008}
Schmitt, F., Kirchmann, P.~S., Bovensiepen, U., Moore, R.~G., Rettig, L.,
  Krenz, M., Chu, J.-H., Ru, N., Perfetti, L., Lu, D.~H., Wolf, M., Fisher,
  I.~R., and Shen, Z.-X., ``Transient electronic structure and melting of a
  charge density wave in {TbTe$_3$},'' {\em Science}~{\bf 321}(5896),
  1649--1652 (2008).

\bibitem{perfetti2006}
Perfetti, L., Loukakos, P., Lisowski, M., Bovensiepen, U., Berger, H.,
  Biermann, S., Cornaglia, P., Georges, A., and Wolf, M., ``Time evolution of
  the electronic structure of {1T-TaS$_2$} through the insulator-metal
  transition,'' {\em Phys. Rev. Lett.}~{\bf 97}(6),  067402 (2006).

\bibitem{hellmann2012time}
Hellmann, S., Rohwer, T., Kall{\"a}ne, M., Hanff, K., Sohrt, C., Stange, A.,
  Carr, A., Murnane, M., Kapteyn, H., Kipp, L., et~al., ``Time-domain
  classification of charge-density-wave insulators,'' {\em Nat. Commun.}~{\bf
  3},  1069 (2012).

\bibitem{Maklar2021_NatCommun}
Maklar, J., Windsor, Y.~W., Nicholson, C.~W., Puppin, M., Walmsley, P.,
  Esposito, V., Porer, M., Rittmann, J., Leuenberger, D., Kubli, M., Savoini,
  M., Abreu, E., Johnson, S.~L., Beaud, P., Ingold, G., Staub, U., Fisher,
  I.~R., Ernstorfer, R., Wolf, M., and Rettig, L., ``Nonequilibrium
  charge-density-wave order beyond the thermal limit,'' {\em Nat. Commun.}~{\bf
  12},  2499 (May 2021).

\bibitem{perfetti2007_PRL_cuprate}
Perfetti, L., Loukakos, P., Lisowski, M., Bovensiepen, U., Eisaki, H., and
  Wolf, M., ``Ultrafast electron relaxation in superconducting
  {Bi$_2$Sr$_2$CaCu$_2$O$_{8+\delta}$} by time-resolved photoelectron
  spectroscopy,'' {\em Phys. Rev. Lett.}~{\bf 99}(19),  197001 (2007).

\bibitem{Cortes2011_PRL_cuprate_dynamics}
Cortés, R., Rettig, L., Yoshida, Y., Eisaki, H., Wolf, M., and Bovensiepen,
  U., ``{Momentum-Resolved Ultrafast Electron Dynamics in Superconducting
  ${\mathrm{Bi}}_{2}{\mathrm{Sr}}_{2}{\mathrm{CaCu}}_{2}{\mathbf{O}}_{8+\ensuremath{\delta}}$},''
  {\em Phys. Rev. Lett.}~{\bf 107}(9),  097002 (2011).

\bibitem{Smallwood2012_BSCCO}
Smallwood, C.~L., Hinton, J.~P., Jozwiak, C., Zhang, W., Koralek, J.~D.,
  Eisaki, H., Lee, D.-H., Orenstein, J., and Lanzara, A., ``Tracking cooper
  pairs in a cuprate superconductor by ultrafast angle-resolved
  photoemission,'' {\em Science}~{\bf 336}(6085),  1137--1139 (2012).

\bibitem{Danz2021_nanoimaging}
Danz, T., Domröse, T., and Ropers, C., ``Ultrafast nanoimaging of the order
  parameter in a structural phase transition,'' {\em Science}~{\bf 371}(6527),
  371--374 (2021).

\bibitem{Johnson2022}
Johnson, A.~S., Perez-Salinas, D., Siddiqui, K.~M., Kim, S., Choi, S.,
  Volckaert, K., Majchrzak, P.~E., Ulstrup, S., Agarwal, N., Hallman, K.,
  Haglund, R.~F., Günther, C.~M., Pfau, B., Eisebitt, S., Backes, D.,
  Maccherozzi, F., Fitzpatrick, A., Dhesi, S.~S., Gargiani, P., Valvidares, M.,
  Artrith, N., de~Groot, F., Choi, H., Jang, D., Katoch, A., Kwon, S., Park,
  S.~H., Kim, H., and Wall, S.~E., ``Ultrafast x-ray imaging of the
  light-induced phase transition in {V}{O}$_2$,'' {\em Nat. Phys.}  (2022).

\bibitem{Schonhense2015_Elsevier}
Schönhense, G., Medjanik, K., and Elmers, H.-J., ``{Space-, time- and
  spin-resolved photoemission},'' {\em J. Electron Spectrosc. Relat.
  Phenom.}~{\bf 200},  94--118 (2015).

\bibitem{puppin2019_RSI}
Puppin, M., Deng, Y., Nicholson, C., Feldl, J., Schr{\"o}ter, N., Vita, H.,
  Kirchmann, P., Monney, C., Rettig, L., Wolf, M., et~al., ``Time-and
  angle-resolved photoemission spectroscopy of solids in the extreme
  ultraviolet at 500 khz repetition rate,'' {\em Rev. Sci. Instrum.}~{\bf
  90}(2),  023104 (2019).

\bibitem{Maklar2020_MetisPhoibos}
Maklar, J., Dong, S., Beaulieu, S., Pincelli, T., Dendzik, M., Windsor, Y.~W.,
  Xian, R.~P., Wolf, M., Ernstorfer, R., and Rettig, L., ``A quantitative
  comparison of time-of-flight momentum microscopes and hemispherical analyzers
  for time- and angle-resolved photoemission spectroscopy experiments,'' {\em
  Rev. Sci. Instrum.}~{\bf 91}(12),  123112 (2020).

\bibitem{ru2006}
Ru, N. and Fisher, I.~R., ``Thermodynamic and transport properties of
  {YTe$_3$}, {LaTe$_3$}, and {CeTe$_3$},'' {\em Phys. Rev. B}~{\bf 73},  033101
  (Jan 2006).

\bibitem{ru2008}
Ru, N., Condron, C.~L., Margulis, G.~Y., Shin, K.~Y., Laverock, J., Dugdale,
  S.~B., Toney, M.~F., and Fisher, I.~R., ``Effect of chemical pressure on the
  charge density wave transition in rare-earth tritellurides {$R$Te$_3$},''
  {\em Phys. Rev. B}~{\bf 77},  035114 (Jan 2008).

\bibitem{Puppin2022_PRB_bandmapping}
Puppin, M., Nicholson, C.~W., Monney, C., Deng, Y., Xian, R.~P., Feldl, J.,
  Dong, S., Dominguez, A., H\"ubener, H., Rubio, A., Wolf, M., Rettig, L., and
  Ernstorfer, R., ``Excited-state band structure mapping,'' {\em Phys. Rev.
  B}~{\bf 105},  075417 (Feb 2022).

\bibitem{brouet2008}
Brouet, V., Yang, W., Zhou, X., Hussain, Z., Moore, R., He, R., Lu, D., Shen,
  Z., Laverock, J., Dugdale, S., et~al., ``Angle-resolved photoemission study
  of the evolution of band structure and charge density wave properties in
  {RTe$_3$} ({R= Y, La, Ce, Sm, Gd, Tb, and Dy}),'' {\em Phys. Rev. B}~{\bf
  77}(23),  235104 (2008).

\bibitem{maschek2015}
Maschek, M., Rosenkranz, S., Heid, R., Said, A.~H., Giraldo-Gallo, P., Fisher,
  I.~R., and Weber, F., ``{Wave-vector-dependent electron-phonon coupling and
  the charge-density-wave transition in TbTe$_3$},'' {\em Phys. Rev. B}~{\bf
  91}(23),  235146 (2015).

\bibitem{laverock2005fermi}
Laverock, J., Dugdale, S., Major, Z., Alam, M., Ru, N., Fisher, I., Santi, G.,
  and Bruno, E., ``Fermi surface nesting and charge-density wave formation in
  rare-earth tritellurides,'' {\em Phys. Rev. B}~{\bf 71}(8),  085114 (2005).

\bibitem{schmitt2011}
Schmitt, F., Kirchmann, P.~S., Bovensiepen, U., Moore, R.~G., Chu, J.-H., Lu,
  D.~H., Rettig, L., Wolf, M., Fisher, I.~R., and Shen, Z.-X., ``{Ultrafast
  electron dynamics in the charge density wave material TbTe$_3$},'' {\em New
  J. Phys.}~{\bf 13},  063022 (Jun 2011).

\bibitem{Maklar2021_PRL}
Maklar, J., Windsor, Y.~W., Nicholson, C.~W., Puppin, M., Walmsley, P.,
  Esposito, V., Porer, M., Rittmann, J., Leuenberger, D., Kubli, M., Savoini,
  M., Abreu, E., Johnson, S.~L., Beaud, P., Ingold, G., Staub, U., Fisher,
  I.~R., Ernstorfer, R., Wolf, M., and Rettig, L., ``Nonequilibrium
  charge-density-wave order beyond the thermal limit,'' {\em Nat. Commun.}~{\bf
  12},  2499 (May 2021).

\bibitem{Wilson1975_TaS2_CDW}
Wilson, J.~A., Di~Salvo, F.~J., and Mahajan, S., ``{Charge-density waves and
  superlattices in the metallic layered transition metal dichalcogenides},''
  {\em Adv. Phys.}~{\bf 24}(2),  117--201 (1975).

\bibitem{Barrett2012}
Barrett, N., Conrad, E., Winkler, K., and Krömker, B., ``Dark field
  photoelectron emission microscopy of micron scale few layer graphene,'' {\em
  Rev. Sci. Instrum.}~{\bf 83}(8),  083706 (2012).

\bibitem{Barrett2013}
Barrett, N., Winkler, K., Krömker, B., and Conrad, E., ``Laboratory-based real
  and reciprocal space imaging of the electronic structure of few layer
  graphene on {S}i{C}(0001¯) using photoelectron emission microscopy,'' {\em
  Ultramic.}~{\bf 130},  94--100 (2013).

\bibitem{Wan2020}
Wan, G., Panditharatne, S., Fox, N.~A., and Cattelan, M., ``Graphene-diamond
  junction photoemission microscopy and electronic interactions,'' {\em Nano
  Express}~{\bf 1},  020011 (jul 2020).

\bibitem{Nicholson2022_TaS2}
Nicholson, C.~W., Petocchi, F., Salzmann, B., Witteveen, C., Rumo, M., Kremer,
  G., von Rohr, F.~O., Werner, P., and Monney, C., ``{Modified interlayer
  stacking and insulator to correlated-metal transition driven by uniaxial
  strain in {1$T$-TaS$_2$}},'' {\em arXiv:2204.05598}  (2022).

\bibitem{Zocco2015}
Zocco, D.~A., Hamlin, J.~J., Grube, K., Chu, J.-H., Kuo, H.-H., Fisher, I.~R.,
  and Maple, M.~B., ``Pressure dependence of the charge-density-wave and
  superconducting states in {G}d{T}e$_{3}$, {T}b{T}e$_{3}$, and
  {D}y{T}e$_{3}$,'' {\em Phys. Rev. B}~{\bf 91},  205114 (May 2015).

\end{thebibliography}

\end{document}